\pdfoutput=1
\documentclass[11pt]{article}

\usepackage{titlesec}
\usepackage{float}

\usepackage{wrapfig}

\titleformat{\section}
  {\normalfont\large\bfseries}
  {\thesection}{.5em}{}
  
\titleformat{\subsection}
  {\normalfont\bfseries}
  {\thesubsection}{.5em}{}
   
  \titleformat{\paragraph}[runin]
  {\normalfont\bfseries}
  {\theparagraph}{0pt}{}

\usepackage[margin=0.97in,top=.99in,bottom=1.1in, textheight=22.51cm,textwidth=16.51cm, heightrounded,columnsep=0.81cm]{geometry}

\usepackage{amsfonts,amsmath}

\usepackage{siunitx}

\pdfoutput=1  %  for arxiv

\usepackage{cite,url}

\usepackage{graphicx}

\usepackage[font={scriptsize}]{caption}
\usepackage[font={scriptsize}]{subcaption}

\usepackage[colorlinks,bookmarksopen,bookmarksnumbered,citecolor=red,urlcolor=red]{hyperref}

\newlength{\noteWidth}
\long\def\notes#1{\ifinner
           {\footnotesize #1}
           \else
           \marginpar{\parbox[t]{\noteWidth}{\raggedright\footnotesize #1}}
       \fi\typeout{#1}}

\def\notes#1{\typeout{read notes: #1}}  %uncomment for final version

\def\spm#1{\notes{SPM:  #1}}

\def\rd#1{{\color{red}#1}}

\def\sfHP{{\text{\sf HP}}}
\def\sfa{{\text{\sf a}}}
\def\sftotal{{\text{\sf total}}}
\def\sfld{{\text{\sf load}}}
\def\sfac{{\text{\sf ac}}}
\def\sfswh{{\text{\sf swh}}}
\def\sffwh{{\text{\sf fwh}}}
\def\sfpl{{\text{\sf pl}}}

\def\sfBP{{\text{\sf BP}}}
\def\sfLP{{\text{\sf LP}}}

\def\power{\varrho}
\def\powertot{\varrho_{{\text{\sf tot}}}}
\def\barpowertot{{\bar\varrho}_{{\text{\sf tot}}}}

\def\rhotr{\rho_{\text{\it tr}}}
\def\tempset{\theta^{\text{\it set}}}

\def\urls#1{{\small \url{#1}}}

\def\generate{{\cal A}}

\def\sq{\hbox{\rlap{$\sqcap$}$\sqcup$}}
\def\qed{\ifmmode\sq\else{\unskip\nobreak\hfil
\penalty50\hskip1em\null\nobreak\hfil\sq
\parfillskip=0pt\finalhyphendemerits=0\endgraf}\fi\medskip}

\long\def\defbox#1{\framebox[.9\hsize][c]{\parbox{.85\hsize}{%
\parindent=0pt
\baselineskip=12pt plus .1pt      % STYLE
\parskip=6pt plus 1.5pt minus 1pt % CHANGES
 #1}}}

\long\def\beginbox#1\endbox{\subsection*{}%
\hbox{\hspace{.05\hsize}\defbox{\medskip#1\bigskip}}%
\subsection*{}}

\def\endbox{}

\def\transpose{{\hbox{\it\tiny T}}}

\newsavebox{\junk}
\savebox{\junk}[1.6mm]{\hbox{$|\!|\!|$}}

\def\argmin{\mathop{\rm arg\, min}}

\def\state{{\sf X}}
\def\stateu{{\sf X}^{\sf u}}
\def\staten{{\sf X}^{\sf n}}

\newcommand{\field}[1]{\mathbb{#1}}

\def\Re{\field{R}}

\def\ind{\field{I}}

\def\bfmath#1{{\mathchoice{\mbox{\boldmath$#1$}}%
{\mbox{\boldmath$#1$}}%
{\mbox{\boldmath$\scriptstyle#1$}}%
{\mbox{\boldmath$\scriptscriptstyle#1$}}}}

\def\bfmm{\bfmath{m}}

\def\bfmD{\bfmath{D}}

\def\bfmU{\bfmath{U}}

\def\bfmX{\bfmath{X}}

\def\tily{\tilde{y}}
\def\tilY{\widetilde{Y}}

\def\tilomega{\widetilde{\omega}}

\def\bfmY{\bfmath{Y}}

\def\bfmhhaY{\bfmath{\hhaY}} %\widehat{\widehat{Y}}}}
\def\bfmhhaY{\hbox to 0pt{$\widehat{\bfmY}$\hss}\widehat{\phantom{\raise 1.25pt\hbox{$\bfmY$}}}}

\def\bfzeta{\bfmath{\zeta}}

\def\til={{\widetilde =}}

 \def\FRAC#1#2#3{\genfrac{}{}{}{#1}{#2}{#3}}

\def\ddt{{\mathchoice{\FRAC{1}{d}{dt}}%
{\FRAC{1}{d}{dt}}%
{\FRAC{3}{d}{dt}}%
{\FRAC{3}{d}{dt}}}}

\def\ddtp{{\mathchoice{\FRAC{1}{d^{\hbox to 2pt{\rm\tiny +\hss}}}{dt}}%
{\FRAC{1}{d^{\hbox to 2pt{\rm\tiny +\hss}}}{dt}}%
{\FRAC{3}{d^{\hbox to 2pt{\rm\tiny +\hss}}}{dt}}%
{\FRAC{3}{d^{\hbox to 2pt{\rm\tiny +\hss}}}{dt}}}}

\def\half{{\mathchoice{\FRAC{1}{1}{2}}%
{\FRAC{1}{1}{2}}%
{\FRAC{3}{1}{2}}%
{\FRAC{3}{1}{2}}}}

\def\eqdef{\mathbin{:=}}

\def\Prob{{\sf P}}

\def\average#1,#2,{{1\over #2} \sum_{#1}^{#2}}

\def\eye(#1){{\bf(#1)}\quad}

\def\Section#1{Section~\ref{#1}}

\def\bary{{\overline {y}}}
\def\barY{{\overline {Y}}}

\newcounter{rmnum}
\newenvironment{romannum}{\begin{list}{{\upshape (\roman{rmnum})}}{\usecounter{rmnum}
\setlength{\leftmargin}{8pt}
\setlength{\rightmargin}{4pt}
\setlength{\itemsep}{1pt}
\setlength{\itemindent}{10pt}
}}{\end{list}}

\newcounter{anum}

\def\bfDelta{\bfmath{\Delta}}

\def\util{\mathchoice{\mbox{\small$\cal U$}}%
{\mbox{\small$\cal U$}}%
{\mbox{$\scriptstyle\cal U$}}%
{\mbox{$\scriptscriptstyle\cal U$}}}

\graphicspath{{figuresHICSS2017/}}

\def\Ebox#1#2{%
\begin{center}
\includegraphics[width= #1\hsize]{#2} \end{center}}

\def\Fig#1{Fig.~\ref{#1}}
\def\Table#1{Table.~\ref{#1}}

\def\ind{\field{I}}

\def\Re{\field{R}}

\title{\vspace{-1cm}
Demand Dispatch with Heterogeneous Intelligent Loads
}

 \author{
 \normalsize
 \begin{tabular}{ccccccc} 
 Joel Mathias & \hspace{0.5cm} & 
 Ana Bu\v{s}i\'c &  \hspace{0.5cm} &	
 Sean Meyn 
 \\ 
Dept.~ECE && Inria Paris && Dept.~ECE 
  \\
 \small
Univ.~of Florida
&&	 	
 \small	
D\'epartement d'Informatique de l'ENS
 && 
 \small   
Univ.~of Florida
   \end{tabular}   
   }
   
     \date{}
   
\begin{document}

\maketitle

\bigskip

\paragraph{Abstract} 

A distributed control architecture is presented that is intended to make a collection of heterogeneous loads appear to the grid operator as a nearly perfect battery.   Local control is based on  randomized decision rules advocated in prior research, and extended in this paper to any load with a discrete number of power states.   Additional linear filtering at the load ensures that the input-output dynamics of the aggregate has a nearly flat input-output response: the behavior of an ideal, multi-GW battery system.
\smallskip{
	
}

\smallskip
{
%\small
\noindent
 \textbf{Acknowledgements}
 {\it Research supported by  NSF grant CPS-1259040, and
 PGMO (Gaspard Monge Program for Optimization and operations research).
% the French National Research Agency grant ANR-12-MONU-0019.
  We thank Rim Kaddah of Telecom Paristech  for many stimulating conversations, especially relating to the design choices discussed in Sections~\ref{s:LC:Markov}
and \ref{s:LC:Inv}
} 
}

\section{Introduction}
\label{s:intro}

\textit{Are billion dollar batteries and billion dollar gas turbine generators  required to manage the volatility of renewable generation?}

  In prior research, it is argued that balancing resources will come from flexible loads at much lower cost and potentially greater performance.  In order to realize this vision, a decentralized control design is utilized; the design respects the limitations of the loads, which are based on dynamic constraints as well as strict bounds on the quality of service (QoS) delivered to consumers \cite{chebusmey14}.   Automation is required to ensure that the grid operator obtains reliable ancillary service, and that reliability to the consumer is also maintained.   

The goal of this  research is to create virtual energy storage from flexible loads.  The framework here and in prior research is \textit{Demand Dispatch}:  power consumption from loads varies in a possibly coordinated manner to automatically and continuously  provide service to the grid, without impacting QoS to the consumer.

This paper is an extended version of the conference paper \cite{matbusmey17} and includes additional details on the design on an optimal inverse filter and the creation of a simulation testbed for demand dispatch. The paper investigates a question posed in \cite{matkadbusmey16}:  \textit{what intelligence is required at the grid-level to implement demand dispatch}?   The question was addressed through extra layers of local control at each load.  One topic left for future research was how to approximately invert the dynamics of an aggregate of loads so that the resulting dynamics would approximate a perfect battery.  
  
%\rd{Sad news in Hawaii:}
%   
%  \urls{http://www.utilitydive.com/news/hawaiian-electric-wants-to-finalize-lng-import-deal-by-q1-2016/409380/}
%  
  
\begin{figure}[!h]
	\Ebox{.75}{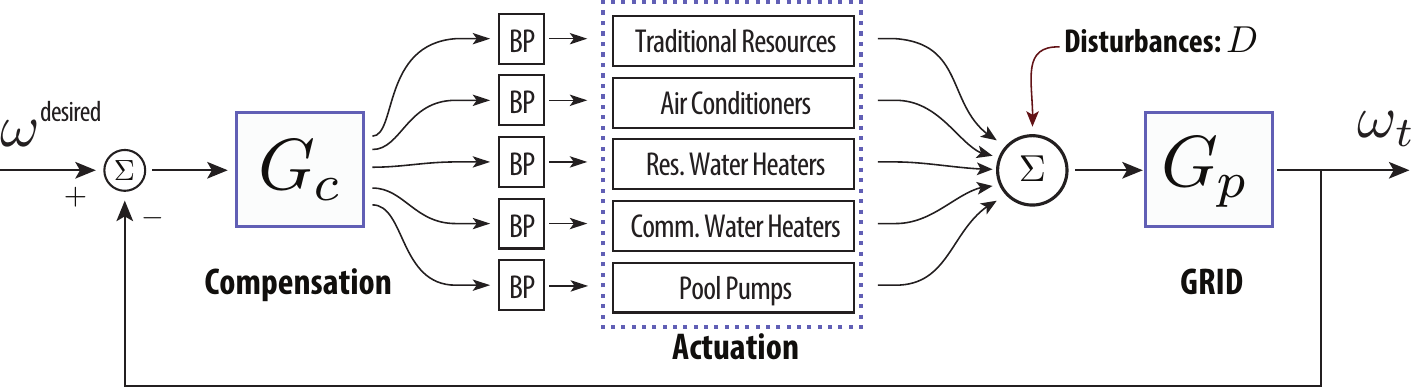}
	\vspace{-1 em}
	\caption{A control architecture for Demand Dispatch.} 
	\vspace{-.75em}
	\label{f:grid}
\end{figure}

The control architecture proposed in \cite{meybarbusyueehr15,matkadbusmey16} is illustrated in \Fig{f:grid}.   The ``compensation'' block represents today's balancing authority (BA),  and the ``grid'' represents the  aggregate dynamics of  loads, generators,   transmission lines, and other grid elements.
Design of the compensator $G_c$ will be based on an input-output model of the grid, denoted $G_p$ in the figure\cite{kun94}  (see \Section{s:GLC} for  details). 
			\spm{Reviewer 1 mentions that Fig 1 is missing generators}
The contributions of this paper are summarized here:
\begin{romannum}
\item  New design techniques are introduced for a broad class of on/off loads (or more generally, loads with a finite number of operating states).   
It is found in numerical results that the resulting mean-field dynamics share desirable properties observed in the optimal design of \cite{meybarbusyueehr15}.  In particular, in every example considered, the linearized dynamics are minimum phase. 

\item  It is argued that the minimum phase property is valuable in the design of a prefilter for each load.   Applying techniques from the theory of robust control, it is shown that the input-output dynamics can be shaped to appear as a constant gain over a bandwidth centered at the nominal period of the load.   

\item  With one-way communication from the BA to the loads, it is shown that the aggregate, with each load acting independently,  serves as a nearly perfect ``virtual battery''.   This is argued based on control analysis, and tested through  simulation.

%Grid-level 
%simulation experiments were conducted using over 50 different classes of loads acting in parallel, 
%with  local control designed for each load.  
\end{romannum}
Grid-level  simulation experiments were conducted using over 50 different types of loads.  Local control ensures that this diverse population --- including pools with time constants of 24 hours, and residential air-conditioning with time constants of tens of minutes to one hour --- act cooperatively to provide regulation over all time-scales.

\paragraph{\textit{Related research}} 
Beginning in the early eighties,
%Following the seminal work of Schweppe,  
deterministic schemes were introduced to model and control a population of  thermostatically controlled loads (TCLs) for ancillary services   \cite{malcho85,cal09}.  Randomized algorithms appeared in the sequels \cite{kocmatcal11,matkoccal13}; system identification and state estimation are required for accurate tracking of the control signal.
 
Centralized control is the subject of \cite{haosanpoovin15}, where the main contribution is to address combinatorial complexity through a priority ordering of loads.   The state information required by the centralized controller presents challenges in terms of both communication and privacy.

% a centralized controller implements a priority-stack-based control algorithm to perform ON/OFF switching decisions for individual TCLs: priority stacks are constructed on the basis of state information (the ON/OFF state and the temperature distance to the deadband limit), and units are turned ON (or OFF) in order of priorities. While the approach avoids short-cycling, 

Local control of refrigerators is proposed for primary frequency control   in \cite{zirvreand15}. 
% following initial ideas of Schweppe
 A randomized control architecture is introduced to avoid synchronization of loads.
 % observed in prior experiments.  
 There are no performance guarantees with respect to ancillary service, which raises concerns  as inaccurate primary frequency control can destabilize the grid \cite{kun94,matkadbusmey16}.

% A randomized control architecture is proposed in which the probability that a load changes power state is a function of the grid frequency deviation.  In addition,  the temperature deadband limits of the loads are shifted dynamically. Lockout times are introduced to mitigate compressor short-cycling.  

 There is substantial literature on indirect load control, where customers are encouraged to shift their electricity usage in response to real-time prices.  Control through price signals can introduce uncertain dynamics
 \spm{this is one conclusion from a very stylized model: that can cause cyclical price fluctuations and increase sensitivity to exogenous disturbances.  }
 and present a risk to system stability \cite{roodahmit12,calhis11,tindjaschustsrb14}.  
 
  \spm{Joel:  conmorbar10  is one of a million papers since Schweppe.  No room for a literature review on real time prices} 

Our approach simultaneously addresses the following four challenges: 
(i) a distributed control architecture simplifies communication infrastructure requirements and assuages consumer privacy concerns; 
(ii) local control ensures reliable ancillary service; 
(iii)   local control also ensures QoS to the consumer; and 
(iv) contractual agreements and periodic credits, such as those proffered by Florida Power and Light in their OnCall program, are advocated to incentivize customer participation.

The present paper concerns all four challenges, but focuses on a new approach to topic (i).   It is assumed that there is only one-way  communication from BA to loads, and that the control signal generated by the grid operator is based on frequency deviation, akin to the manner in which AGC (automatic generation control) is synthesized today.    

It is remarkable to see the potential for demand dispatch based on minimal communication.  However,   some 
communication from loads to the grid operator remains valuable in practice.   In particular,  the grid operator requires bounds on the capacity of service from loads, and may want to occasionally update or verify parameters in local control algorithms.  
 
 The remainder of this paper is organized as follows. \Section{s:DCA} provides details of the distributed control architecture, including grid-level control, local control design, and load dynamics. In \Section{s:sim}, multiple simulations are performed to demonstrate the validity and utility of demand dispatch. Conclusions and directions for future research are summarized in \Section{s:conc}.

\section{Distributed Control Architecture}
\label{s:DCA}

\subsection{Grid level control}
\label{s:GLC}

\begin{figure}
	\Ebox{.6}{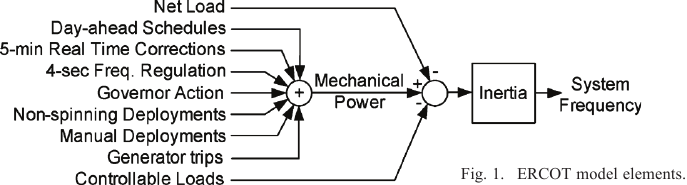}
	\vspace{-1 em}
	\caption{Macro grid model of \cite{chabalsha12}.} 
	\vspace{-.75em}
	\label{f:fig1ERCOT-Grid}
\end{figure}

The macro grid model used in this study is taken from \cite{chabalsha12}, which is itself based on standard power systems analysis  \cite{kun94}.  The grid is modeled as an input-output linear system whose input is power deviation and output, frequency deviation. The elements of their model are illustrated in   \Fig{f:fig1ERCOT-Grid}, which is taken from  Fig.~1 of \cite{chabalsha12}.  A particular  grid model of \cite{chabalsha12} is used in  numerical experiments: 
\begin{equation}
%G_p(s) =
%   \frac{ 0.644 s + 0.147}
%   {
% s^2 + 0.4797 s + 0.147}
%The transfer function from AGC to the output frequency deviation for the ERCOT grid:
 G_p(s) =
 10^{-5}
   \frac{ 2.488 s + 2.057}
   {
 s^2 + 0.3827  s + 0.1071}
\label{e:ERCOTex1}
\end{equation}
The impulse response of this system is in close agreement with the response of frequency to a grid outage in the ERCOT region --- a full discussion can be found in \cite{matkadbusmey16}.

Throughout the paper, the transfer function \eqref{e:ERCOTex1}
 is used to model the ``GRID'' shown in  \Fig{f:grid}.     The resonance of this transfer function corresponds to time-scales on the order of seconds, while in this paper the relevant disturbances to be rejected evolve on much slower time-scales.  This justifies the use of PI control for the choice of $G_c$ in the ``compensation block''.  The output of the compensator is denoted
\[
U_t =  K_P \tilomega_t  + K_I \int_0^t \tilomega_r \, dr
\]
where $\tilomega_t = \omega^{\text{\scriptsize desired}} - \omega_t$.    
The interpretation of $U_t$ is the desired change in power from all resources (MWs).  The control parameters (proportional and integral gains $K_P$, $K_I$)  are chosen to respect the uncertainty of grid dynamics on timescales of seconds or faster.    

In practice, the PI compensator would be modified to avoid ``integrator windup'', as is common practice in synthesizing the AGC signal today.   

For simplicity, in this paper, we focus solely on techniques for balancing within the control region (this includes ramp services, balancing reserves, and  frequency regulation).   In practice, the regulation of tie-line error would be performed in conjunction with these services.
\spm{revised in Sept. to emphasize the breadth of services}

The signal $\bfmU$ is decomposed using several bandpass filters:
Each block ``BP'' shown in \Fig{f:grid} represents a bandpass filter that is chosen based on the dynamics and constraints of the associated aggregate of resources.  Batteries and flywheels are valuable for the highest frequency component of $\bfmU$;  demand dispatch based on refrigerators and water heaters can provide service on time-scales of tens of minutes to several hours   \cite{meybarbusyueehr15}.

\subsection{Local control:  Markovian dynamics}
\label{s:LC:Markov}

\spm{    exponential   sampling far more elegant.  We'll deal with geometric sampling somehow!}  

The local control described here is a continuous-time variant of the myopic design introduced in  \cite{busmey16v}.    

The starting point is the construction of a Markovian model for nominal behavior of an individual load.    The state process evolves in continuous time, on a finite state space denoted $\state$. Hence its dynamics are defined by a \textit{rate matrix}, denoted $\generate_0$.   For two states $x,x'\in\state $ the transition probability is denoted $P^t(x,x') = \Prob\{X_t=x'\mid X_0=x\}$,  which is   the matrix exponential $P^t = \exp(t \generate_0)$,  $t\ge 0$.

It is assumed that the nominal model has a unique invariant pmf (probability mass function), denoted $\pi_0$.  Invariance requires that $\sum_x  \pi_0(x) \generate_0(x,x') =0$ for every $x'\in\state$.

The rate matrix is assumed to be of the following form,
\begin{equation}
\generate_0 =    r [ -I + S_0 ]
\label{e:rate}
\end{equation}
where $S_0$ is a Markov transition matrix, and $I$ is the identity matrix.  
A Markov process $\bfmX$ with rate matrix \eqref{e:rate}
can be realized by first constructing a Poisson process with rate $r$ and jump times $\{T_k : k\ge 1\}$.
The continuous-time process $\bfmX$ is constant on the inter-jump time-intervals $[T_k,T_{k+1})$, and  
\[
\Prob\{ X_{T_{k+1}} = x' \mid X_{T_k} = x\} = S_0(x,x')\,,
\] 
for $  x,x'\in\state$ and $k\ge 0$, with $T_0=0$. The assumption that $X_t=X_{T_k}$ for $t\in [T_k, T_{k+1})$ reflects the fact that we are only considering the load at the sampling times $\{T_k\}$.   \textit{We do not assume that the load state itself is constant over this period.  }

  \spm{no room:   $\Delta_k= T_k-T_{k-1}$ for $k\ge 1$, it follows that $\bfDelta =\{\Delta_k\}$ is i.i.d., with common mean $1/r$.   
}

The state of the load has the following form: $X_t = (X_t^u, X_t^n)$ for $t\ge 0$, and  the state space has the form $ \state=\stateu\times\staten$.  The first component $\bfmX^u$ represents a variable that can be adjusted directly, such as   power consumption, or the temperature set-point for a refrigerator.     The second component $\bfmX^n$ is \textit{indirectly} controlled through  $\bfmX^u$ and exogenous disturbances (e.g.,  someone opens the refrigerator).   

The Markovian  dynamics  for the nominal model are assumed to be of the form
\begin{equation}
S_0(x,x') = R_0(x, x_u') Q_0(x,x_n') ,
\label{e:PQ0R0}
\end{equation}
where $x,x'\in \state=\stateu\times\staten$, and $\sum_{x_u'} R_0(x, x_u') = \sum_{x_n'} Q_0(x,x_n') = 1$.   The matrices $R_0$, 
$Q_0$ model the dynamics of $\bfmX^u$, $\bfmX^n$, respectively.
\spm{Reviewer 1 asks the meaning and significance of $R_0$ and $Q_0$. }

  The construction of $S_0$ is of course entirely dependent on the characteristics of the particular load.  

For simplicity,  in this paper it is assumed that $X^u_t =m_t$ represents power consumption (that can be controlled directly at the load).  It is assumed moreover that there are  just two power states:  on or off.     
The process $\bfmm$ evolves in the binary set denoted
$\stateu=\{\ominus,\oplus\}$.  Denote by $  \util(x)$ the associated power consumption:  $  \util(\ominus, x_n) =0$, and $  \util(\oplus, x_n) = \power$~(kW) (a positive value, independent of $x_n\in \staten$).

Local control is based on a perturbation of nominal behavior, defined by a family of rate matrices $\{\generate_\zeta : \zeta\in\Re\}$.   The following myopic design is used in all of the numerical experiments considered here:
\begin{equation}
\!\!
S_\zeta(x,x') = S_0(x,x') \exp(\zeta \util(x') - \Lambda_\zeta(x))
\label{e:PQ0R0myopic}
\end{equation}
in which $\Lambda_\zeta(x)$ is the normalizing constant defined so that $\sum_{x'} S_\zeta(x,x')=1$ for each $x$.   

%This and alternate designs are introduced in \cite{busmey16v}.    

The goal of the myopic design is to influence the load to consume more power at time $t$ when $\zeta_t>0$, and less power when $\zeta_t<0$.

Given a homogeneous collection of $N$ loads, the empirical distribution at time $t$ is defined as follows:
\begin{equation}
\mu^N_t(x)\eqdef \frac{1}{N}\sum_{i=1}^N  \ind\{ X^i_t =x \} ,\quad x\in\state.
\label{e:empDist}
\end{equation}
We assume this is approximated by the mean-field equations,
\begin{equation}
\ddt \mu_t  = \mu_t \generate_{\zeta_t} 
\label{e:ctsMFM}
\end{equation}
in which $\mu_t$ is interpreted as a row vector;  justification for large $N$ is straightforward in the discrete-time setting \cite{meybarbusyueehr15}.   The average power is denoted $y_t = \sum_x \mu_t(x) \util(x)$, and the steady-state average power consumption for the nominal model is $\bary^0 = \sum_x \pi_0(x) \util(x)$.

It is assumed moreover that $\generate_\zeta$ is continuously differentiable in $\zeta$.  
This justifies the linear state space model approximation,
\begin{equation}
\begin{aligned}
\ddt
 \Phi_t = A \Phi_t + B \zeta_t\,,
 \quad
\gamma_t = C \Phi_t
\end{aligned}
\label{e:LSSmfg}
\end{equation}
where
$A=\generate^\transpose_0$, and    
$B$,
$C^\transpose$ are column vectors of dimension $d=|\state|$:
\[
C_k = \util(x^k) \,, \ \  B_k = \sum_x\pi_0(x) \generate'_0(x,x^k) \,, \quad 1\le k\le d
\]
where $\generate'_0$ is the derivative of $\generate_\zeta$ at $\zeta=0$. 

The state  $\Phi_t$ is $d$-dimensional:  $\Phi_t(k)$ is intended to approximate $\mu_t(x^k) -\pi_0(x^k)$ 
for $1\le k \le d$.   The output $\gamma_t$ is an approximation of $\tily_t = y_t - \bary^0$.

\spm{subscripts here and $_t $ later!  Need to work this out}

\subsection{Local control:  inverse filter design}
\label{s:LC:Inv}

\begin{figure}[!h]
	\Ebox{.75}{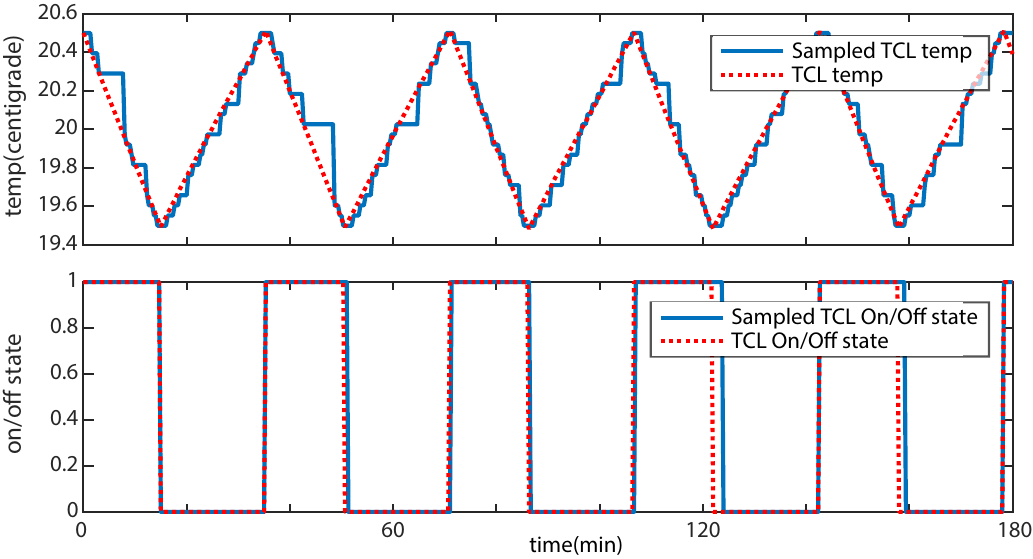}
	\vspace{-.75em}
	\caption{Temperature evolution of a TCL.} 
	\label{f:tempvtime1}
	\vspace{-.3cm}
\end{figure}

\Fig{f:tempvtime1} shows the nominal behavior of an air-conditioning load along with the associated Markovian model, whose sample paths are piece-wise constant.   The sampling rate $r$ was chosen so that the mean sampling time $1/r$ is much smaller than the nominal period of the load.  

Consider a collection of 2,000 similar units, each consuming 1kW  of power when operating, and 50\%\ duty cycle.  Hence,  without any coordination, the average power consumption is about $\barY^0\eqdef$~1MW.   

Denote the total power consumption at time $t$ by $Y_t$.   Equivalently,
\[
Y_t = N \sum_x \mu_t^N (x) \util(x)\,,
\]
and denote the deviation $\tilY_t= Y_t - \barY^0$.   Using centralized control, the loads could be coordinated so that  $\tilY_t$  tracks a square wave of amplitude 1MW  nearly perfectly.   If the frequency of this square wave is chosen to be  the nominal period (approximately 30 mins in this example), then each load would appear to be evolving without external influence.  If the frequency is far from the nominal, then the load will receive poor QoS:  either excessive cycling, or poor temperature control.  
The decentralized control strategy described here is designed to respect these constraints.   

\begin{figure}[h]
\Ebox{.75}{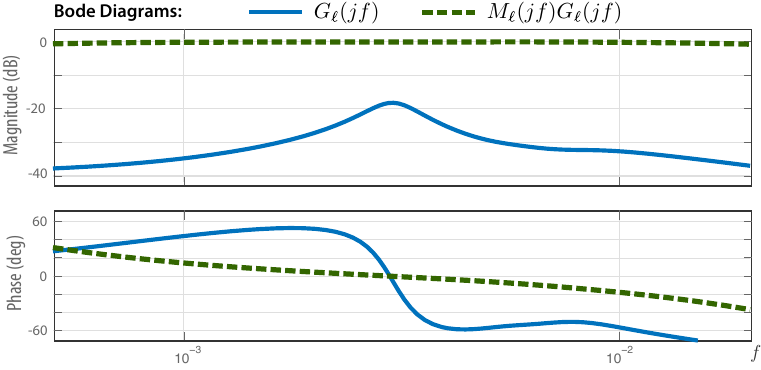}
%iiTCL-noM.pdf}
	\vspace{-.75em}
\caption{Linearized mean-field dynamics  for a TCL with and without inverse-filter}
\label{f:iiTCL}
	\vspace{-.75em}
\end{figure}

In every design considered, it is found that the aggregate dynamics exhibit a  resonance near the nominal frequency.    
\Fig{f:iiTCL} shows a Bode plot for a linearized TCL model with transfer function $G_\ell$, with resonance  at  $f_r=3\times 10^{-3}$~rads/sec (consistent with a 30 min period).
\spm{  \rd{  This plot corresponds to an AC with a 35 minute cycle time and a 43\%\ duty cycle.}
I am tempted to redefine the myopic design, $S_\zeta(x,x') = S_0(x,x') \exp(\zeta g \util(x') - \Lambda_\zeta(x))
$  where  $g>0$ is a fixed scaling factor.   This way I can increase the magnitude on your bode plot ($g=10$ means I can increase by 20dB)}

These observations are motivation for restricting the bandwidth of service from each load to a neighborhood of this resonance, and introducing pre-filtering to flatten the resonance.

%\begin{wrapfigure}{r}{.4\hsize}
%\Ebox{.75}{iiTCL-noM.pdf}
%\caption{Linearized mean-field dynamics for a TCL with and without inverse-filter.} 
%\end{wrapfigure}

For each load, an associated mean-field model and its linearization can be computed exactly.   In each example that we have considered, these dynamics are minimum phase, which simplifies the \textit{inverse filter design} proposed here. The outcome of this design is a prefilter that removes the resonance, and makes the linearized dynamics appear all-pass within a prescribed bandwidth.

Let $G_\ell$ denote the transfer function for the linearized mean-field model,  and let $M_\ell$ denote the pre-filter.   The goal is to design the pre-filter so that $M_\ell(jf) G_\ell(jf) \approx 1$ for a range of $f\in\Re$.  This goal can be re-cast as the robust control problem described next.

 \begin{figure}[!h]
\Ebox{.5}{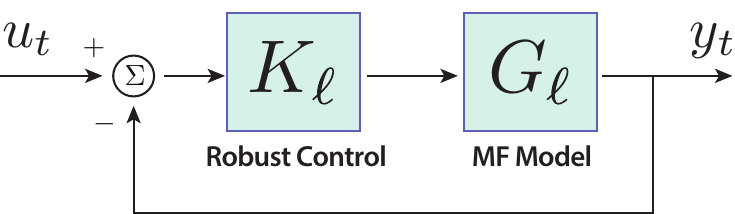}
 \vspace{-.75em}
\caption{Feedback control system for Inverse Filter Design.} 
\label{f:localFilterDesign}
\vspace{-.3cm}
\label{f:fbi}
\end{figure}

\Fig{f:localFilterDesign} shows a  feedback control loop in which $K_\ell$ is a transfer function to be designed.  In the standard robust control framework, one objective for design is to ensure that the transfer function from $\zeta_t$ to $y_t$ is nearly unity in some frequency range.   This closed loop transfer function is equal to $L_\ell/(1+L_\ell)$, where $L_\ell = K_\ell G_\ell $.    Consequently, a solution to the robust control problem provides an inverse filter design solution    $M_\ell = K_\ell/(1+L_\ell)$.   The desired approximation $M_\ell G_\ell =L_\ell/(1+L_\ell)\approx 1$ is obtained in the specified frequency band. 
% is called the ``loop transfer function''. 

%\begin{wrapfigure}{r}{.4\hsize}
%\Ebox{.75}{iiTCL-noM.pdf}
%\caption{Linearized mean-field dynamics for a TCL with and without inverse-filter.} 
%\end{wrapfigure}

The robust control problem is posed as an optimization problem over transfer functions. 
The optimal transfer function $K_\ell$ is obtained numerically in Matlab using the {\tt mixsyn} command  \cite{kwa93,bal05}.

\spm{  \rd{ : it would be good to explain this as we have space left. Need to only say that we assign weights $W_1, W_2,W_3$ corresponding the sensitivity, noise sensitivity, and complementary sensitivity transfer functions, respectively. Define q(K). $K_l = \argmin_K q(k)$. The weights are chosen according to the control bandwidth requirements specific to the class of loads. All of this is here: \url{https://www.dropbox.com/sh/9c7e9m3mbrkmz7r/AACMwQ_jHKfi_zGRVdXtUKOya?dl=0}  }}

\subsection{Design with heterogeneous loads}
\label{s:LC:Hetero}

We conclude this section with a few details required to incorporate multiple heterogeneous loads in the demand dispatch model.  

First, observe that the inverse design whose linearization is plotted in
\Fig{f:iiTCL} may result in a local control algorithm that is too aggressive for a TCL load.  
\spm{Not clear ``this load'' .   We don't even know what load it is! }
Without the inverse filter, the gain from an aggregate of these TCLs falls quickly for frequencies $f>f_r$, which suggests a problem with this inverse filter design:  excessive cycling of individual loads will occur if the aggregate tracks high frequencies with significant magnitude.   On the other hand, capacity of low-frequency tracking is small because of the temperature constraints associated with TCL hysteresis.

Therefore, it is essential to introduce a second filter to restrict the bandwidth to a range appropriate for the corresponding class of loads.     Specifics are provided in the experimental results surveyed in the next section.   

\smallbreak

This section is concluded with a brief description of the nominal behavior of a TCL, and a summary of the parameters  used in this paper when considering a collection of heterogeneous loads.

A common model for temperature evolution  is the first order differential equation, 
\begin{equation}
\ddt \Theta_t  = -  \frac{1}{RC} (\Theta_t  - \Theta_t^a  +  \theta^g m_t)  +   W_t ,
\label{e:tcltemp}
\end{equation}
in which $\Theta_t $ is the   internal temperature, $\Theta_t^a $ is   ambient temperature,  $C$ is  thermal capacitance, $R$ is   thermal resistance, and $W_t  $ models disturbances.   

The temperature gain parameter is $\theta^g = R \rhotr$, where $\rhotr$ is   the   energy transfer rate:   $\rhotr$ is positive for  TCLs providing cooling,  and negative otherwise.  The power consumption is the ratio $\power = |\rhotr| /\text{COP}$, where the denominator is known as the \textit{coefficient of performance}. 
\spm{Joel:  Power is Positive!!}
The nominal behavior is defined by a   temperature set-point $\tempset$ and a  dead-band range $\delta$,  so that $\Theta_t \in [ \tempset-\delta/2,\tempset+\delta/2]$.   Temperature is regulated to this band via the binary-valued process $\{m_t\}$,  whose behavior is defined by  hysteresis, as illustrated in \Fig{f:tempvtime1}.

%\spm{use 1En notation?}
% https://en.wikipedia.org/wiki/Power_of_10

\begin{table}[t]
	\centering
	{\small 
	\begin{tabular}{||c  c  c c||} 
		Par.  & AC & Fast WH & Slow WH 
		\\ [0.5ex]  
		$\tempset$ &  18--22 & 48--52 & 48--52\\  %temp.\ set-point ($C$) &
		$\delta$ & 0.8--1 & 2.95--3 & 3.95--4 \\  % temp.\ deadband range ($C$) &
		$\Theta^a$ &  30--34 & 19--21 & 19--21 \\  %ambient temp.\ ($C$) &
		$RC$ &  3.5--4.5 & 30--36 & 67--73 \\  %thermal time const.\ (hrs) &
		$\power$ &   14/2.5 & 5/1 & 5/1 \\ [.5ex]  % tr.\ rate ($kW$) / coeff.\ of perf.  &  
	\end{tabular}
	}
	\caption{TCL parameters for ACs and   water heaters.}
	\vspace{-1em}
	\label{table:tclparameters}
\end{table}

\Table{table:tclparameters} displays the ranges of values of  TCL parameters for air-conditioners and electric water heaters (a subset of those surveyed in \cite{matdyscalros12}). The value of $RC$ is in units of time (hrs). The last row denotes the maximal power consumption, $\power = |\rhotr| /\text{COP}$ (kW).

The  experiments that follow are based on a heterogeneous collection of loads in which the parameters for the TCLs take on  values within these limits.  
% that characterize \eqref{e:tcltemp}. 

\section{Simulating the Grid}
\label{s:sim}

It is found in prior numerical studies that the mean field model accurately matches the dynamics of an aggregate of loads,   provided the total number of loads engaged is on the order of hundreds or more \cite{chebusmey15b,meybarbusyueehr15}. These prior works focused primarily on residential pool pumps;  extensions to TCLs are treated in  \cite{chebusmey15b},  but without any supporting simulations.

Simulations demonstrating the tracking and disturbance-rejection performance along with a cost analysis of demand dispatch are presented in this section.  The impact of daily periodic patterns of response from loads is also investigated.   

\spm{text may be useful, but claims may not be fully justified:
 ... demonstrate that the mean-field model tracks the aggregate for a broader range of loads.  The impact of filtering is also investigated.  It is found that the resulting non-linear dynamics are nearly all-pass over a certain bandwidth,
even though  the pre-filter is designed based on the linearized model, without regard to the full nonlinear dynamics of the mean field model.
}

\subsection{Design of a virtual battery}
\label{s:DesignBattery}

%To model the behavior shown in \Fig{}, we set $r=1/60~s^{-1}$ for ACs. Fast EWHs are on for $15$ minutes and off for $2.5$ hours, while slow EWHs remain on for $25$ minutes and off for $9 hours$, approximately. To model this asymmetry in on-off durations, we set $r_{\mathrm{on}} = 1/40 ~s^{-1}$  and $r_{\mathrm{off}} = 1/500~s^{-1}$ for fast water heaters, and $r_{\mathrm{on}} = 1/40 ~s^{-1}$  and $r_{\mathrm{off}} = 1/1000 ~s^{-1}$ for slow water heaters. We quantize the TCL temperature interval into $d/2 = 20$. Consequently, $|\mathsf{X}| = d = 40$. 

The experiments conducted involved four classes of loads:   residential air conditioners (AC); small electric water heaters with faster cycle times (f-WH);   large electric water heaters with slower cycle times (s-WH); and residential pool pumps.   They are distinguished by their nominal period, and based on this, a bandwidth of service was chosen for the design of each bandpass filter.  In each case, a second-order Butterworth filter was adopted --- the parameters are summarized in \Table{t:loadchar}.

  \begin{figure}[h]
	\Ebox{.9}{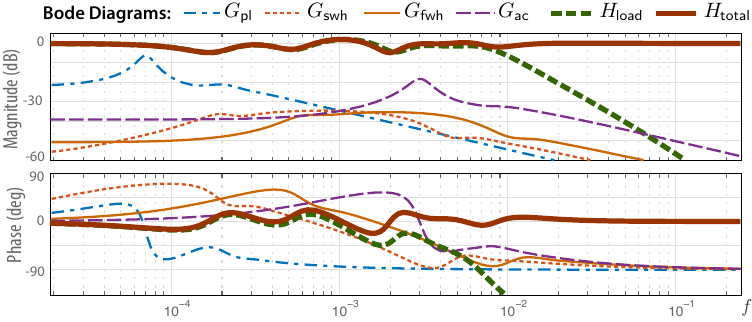}
	\vspace{-1em}
	\caption{Individual and aggregate load dynamics.} 
	\vspace{-.1em}
	\label{f:DDBode}
\end{figure}

Twenty different subgroups were obtained for each TCL class,  through uniform sampling of the values in \Table{table:tclparameters}. Each subgroup contains 2,000 loads, implying a total of 40,000 loads in each TCL class.   The Markovian model was obtained via Monte-Carlo based on simulations of (\ref{e:tcltemp}), following  Section~IV of \cite{busmey16v} and prior work.  The experiments include 40,000 homogeneous pools with 12 hour cleaning cycles, with a nominal Markov model identical to that used in \cite{chebusmey15b}.

 \begin{figure}
 	\Ebox{1}{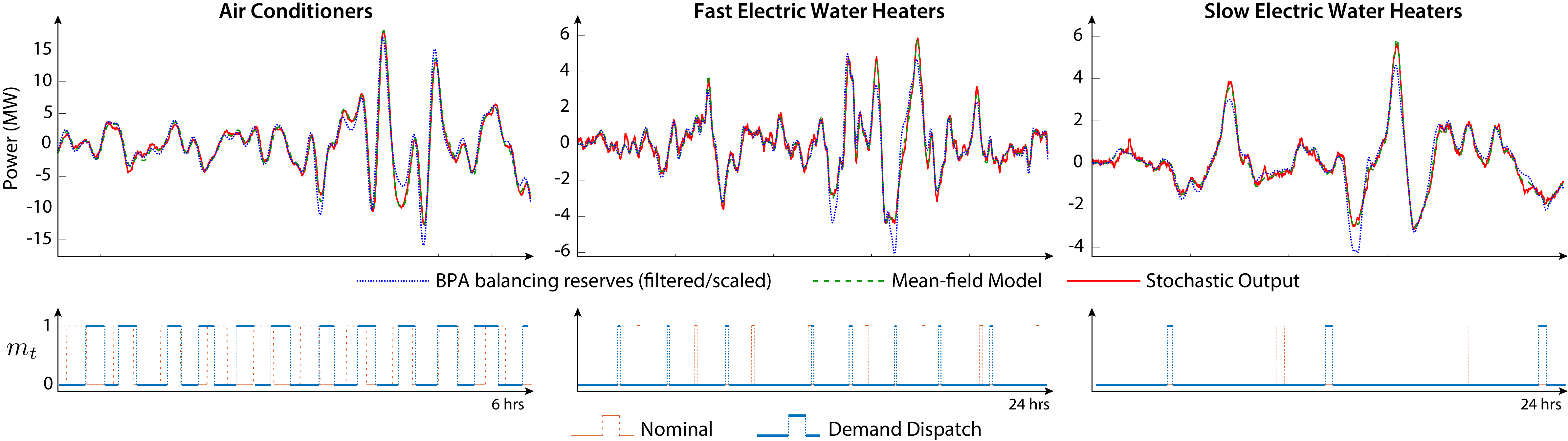}
 	\vspace{-1 em}
 	\caption{Open-loop tracking with 40,000 heterogeneous TCLs.  The lower plots show the on/off state $\bfmm$ for a typical load.} 
 	\vspace{-.3em}
 	\label{f:trackingI}
 \end{figure} 

For each homogeneous subgroup, the controlled Markov model $\{ S_\zeta: \zeta\in\Re\}$ was obtained using the myopic design described in \Section{s:LC:Markov}.    Based on the resulting
design, its linearized dynamics were obtained about $\zeta\equiv 0$.  This was the basis of the inverse filter design   described in \Section{s:LC:Inv}.  In addition, as  described in \Section{s:LC:Hetero}, each load locally pre-filters the regulation signal using a  bandpass filter.

\spm{I feel uncomfortable with LPF for pools.  Changed to this: $[1/24, \, 1/3]$ for now. 
Note that I modified the Bode plot -- removing frequencies below $1e-5$ }

Denote the respective transfer functions for the linearized mean-field model by, respectively,     $G_{\sfac}$,
 $G_{\sffwh}$,  $G_{\sfswh} $,  $G_{\sfpl}$,   and the respective filters (inverse $\times$ band-pass) by   
  $M_{\sfac}$, $M_{\sffwh}$,  $M_{\sfswh} $,  $M_{\sfpl}$.    The linear model of the aggregate dynamics of all the loads is defined by the sum,
\begin{equation}
\!\!\!
H_{\sfld} = M_{\sfac}G_{\sfac} + M_{\sffwh}G_{\sffwh} + M_{\sfswh}G_{\sfswh} + M_{\sfpl}G_{\sfpl}
\label{e:ld}
\end{equation}
The Bode plot for $H_{\sfld}$ is shown in 	\Fig{f:DDBode}.    The rapid decline in the 
magnitude plot beyond $f=10^{-2}$~rads/sec is due to the inherent bandwidth constraints of the loads.
Hence, the actuation is augmented with an ideal resource $G_{\sfa} \equiv 1$.   A high pass
filter $M^{\sfHP}$ was designed with bandwidth beyond $f=10^{-2}$ rads/sec, 
so that the introduction of this resource flattens out the Bode plot. The total response is
modeled by the transfer function
$H_{\sftotal} = H_{\sfld} + M^{\sfHP}G_{\sfa} $,
 whose Bode plot is also shown in 	\Fig{f:DDBode}. 

The actuation obtained from $G_{\sfa}$ might come from  batteries, responsive generators,  or fast responding loads that provide accurate tracking.   The time-scales of ancillary service from these resources are assumed to be in the range of primary control, which is why accurate response is needed in this bandwidth.  
\spm{following comment by reviewer 2 needs to be addressed: "The  inclusion of an “ideal resource” needs more discussion}
% Joel starts : added || for avg power and total power because heating loads have a -ve power convention

\begin{table}[h]
\medskip

 	\centering
	
%	multiply by (60^2)/(2*pi)
%	{\small 
%	\begin{tabular}{||c c c  c||}  
%	Load & Period & filter BW (rad/s)  & Avg. pwr. 
%		\\ [0.5ex]  
%	AC & one hr.  &   $[2\ttimes10^{-3}, 8\ttimes10^{-3}]$&  \rd{500 kW !!}
%	\\
%	f-WH  &     2--4 hrs.  &  $[6\ttimes10^{-4}, 2\ttimes10^{-3}]$&  500 kW
%	\\ 
% s-WH  &  8--10 hrs.  &  $[2\ttimes10^{-4}, 2\ttimes10^{-3}]$&  500 kW
% \\
% Pools & 24 hrs.  &  $[0, 5\ttimes10^{-4}]$  &  500 kW
%	\end{tabular}
%}

	{\small 
	\begin{tabular}{||c c c  c c||}  
	Load & Period &  BW (cyc/hr) & $  \barpowertot  $   & $ \powertot  $
		\\ [0.5ex]  
	AC & 20min--1hr.  &   $[1,\, 1/0.2] $& 97 & 224
	\\
	f-WH  &     2--4 hrs.  &  $[ 1/3 , \, 1/0.5]$&11   & 200
	\\ 
 s-WH  &  8--12 hrs.  &  $[1/9,\,  1]$&  8.5   & 200
 \\
 Pools & 24 hrs.  &  $[1/24, \, 1/3]$  &  20  & 40 
	\end{tabular}
}
	\caption{Load Dynamics and Power Characteristics:  BW of    $M^{\sfBP}$;
	max    and average power $\powertot$, $\barpowertot$  in MW for 40,000 % $4\times10^{4}$ 
	loads.}
	\label{t:loadchar}
\end{table}

% Joel ends

Have we constructed a perfect battery?   Recall that the nonlinear dynamics have been linearized for the sake of analysis,  but the aggregate dynamics remain nonlinear.  Moreover,  the Bode plot for the linearized dynamics with transfer function $H_{\sftotal}$ is not entirely flat in magnitude or phase.   These shortcomings are no different than what would be expected for a generator providing balancing service, or  a realistic (and imperfect) battery system.   

The next results illustrate the accuracy of tracking, and the application of the ensemble of loads for balancing the grid.

\subsection{Open loop tracking}

The balancing reserves deployed (BRD) from the Bonneville Power Administration (BPA)  were used as a reference signal to evaluate open-loop tracking.   A single typical windy day,  February 19, 2016, was chosen in the open-loop experiments described here.  These experiments illustrate the input-output behavior of each collection of TCLs. 

 For each of the three classes of TCLs, the BRD were passed through a bandpass filter designed based on the frequency characteristics of the class.   \Fig{f:trackingI} shows the \textit{open loop} tracking performance in each case (for the case of AC, the plot shows only six hours during the day).  The tracking accuracy is remarkable for a one-way communication architecture from the grid operator to the loads.
 
We estimate that the AC trajectory represents only 20\%\ of capacity (the signal could be scaled up by  5 while maintaining reasonable tracking), and the other two plots represent about 50\%\ of capacity.     While
demand dispatch does increase cycling of TCLs,  in these experiments, it was found that cycling was increased by only about 5\%\ from nominal.  Without the inclusion of ``opt out'' control,  additional cycling will increase as the magnitude of the reference signal increases \cite{chebusmey14}.   

% Full details will appear in a journal version of this paper.

\spm{For journal paper, discuss capacity based on commented text below.}

%
%Below are the average duty cycles for each category of loads. Since the tracking experiments involved uniform sampling to generate 20 different subgroups for each category, the subgroups to calculate these results will differ slightly from the ones used to generate the tracking figures of the paper.
%
%
%Average duty cycles: (i) 43.27% (AC); (ii) 5.61 % (f-wh); (iii) 4.23% (s-wh)
%
%
%Average power for a single load: (i) 2.423 kW (AC); (ii) 0.28 kW (f-wh); (iii) 0.21 kW (s-wh)
%
%
%Experiments were for 40,000 loads
%
%Capacities:
%
%1. AC --- 40000 * min(3.177, 2.423) / 1000 = 96.92 MW
%
%2. f-wh --- 40000 * min(4.72, 0.28) / 1000 = 11.2 MW
%
%3. s-wh --- 40000 * min(4.79, 0.21) /1000 = 8.4 MW
%
%
%In Fig. 6: (i) the ACs are tracking a signal between +/- 15 MW; (ii) f-wh are tracking a signal approx between +/- 5 MW; (iii) s-wh are tracking a signal approx between +/- 4.5 MW.
%
%
%Please let me know in case any additional information is required.
% 

The entire BRD signal can be tracked   using a combination of pools and the three classes of heterogeneous TCLs along with the high-frequency ideal resources $M^{\sfHP}G_{\sfa}$.   
Results from experiments in non-ideal settings are described next.

% {KirbyTrendsPoorTracking.pdf}

%Figures:
%
%7.    (Section IIIA. Open loop tracking) heterogeneous AC, Fast WH, Slow WH, Pools
%
%8.    (Section IIIA. Open loop tracking) AC + Fast WH + Slow WH + Pools + “perfect actuators”

\subsection{Closed loop performance}
\label{s:cl}

Simulations  were performed to evaluate the disturbance rejection performance of the demand dispatch control architecture.  The experiments were based on the closed loop system represented in \Fig{f:grid}.  

The simulations were run in continuous time using Simulink. Full details are contained in the Appendix.

The nonlinear mean field model tracks the aggregate of loads perfectly in all cases considered.  In particular,
in each of the simulation results shown in \Fig{f:trackingI},   the mean-field model output is nearly indistinguishable from the aggregate stochastic  output. 

Since it is much faster to simulate the nonlinear deterministic system, we see no reason to conduct a stochastic simulation in these experiments.

The demand dispatch simulation model was based on 1 million ACs, 5 million f-wh, 
 5 million s-wh,
and a large number of pools (this number was taken as a parameter in this study). 
Each group of loads evolves according to the corresponding mean-field model \eqref{e:ctsMFM}, which is linear in the state and nonlinear in the input.   

For a homogeneous group $\ell$, there is by design a controlled generator $\{\generate_\zeta^\ell\}$, and a linear filter $M^\ell$ that determine local control.  The dynamics of the aggregate of loads in this subgroup evolves as   
\[
\ddt \mu_t^\ell  = \mu_t^\ell \generate^\ell_{\zeta^\ell_t} ,\quad \zeta^\ell_t = M_\ell \zeta_t\,  \qquad t\ge 0.
\]

Given the specified mix of TCL loads and assuming that the closed-loop system is driven by the BRD signal as the disturbance entering the grid, in order to obtain a flat Bode plot for total actuation as shown in 
\Fig{f:DDBode}, we would require \textit{4 million pools}!     With a PI control architecture, a flat response at low frequencies is not necessary, so experiments were conducted with 1 million residential pools (the approximate number of pools in Florida).   The maximum load is thus 1GW, and the average load is 500~MW,  so that the pools can at best track signals of $\pm 500$~MW.  Tracking was poor when  the BRD signal exceeded this range.

Other resources such as commercial water chillers could be added to increase capacity at low frequencies.  Instead, in the next set of simulations, the pools were augmented with a single 1~GW generator.  This was modeled through the introduction of an additional ideal actuator:
\[
\textstyle
H_{\sftotal} = H_{\sfld} + M^{\sfHP}G_{\sfa} + \frac{1}{4} M^{\sfLP}G_{\sfa}
\]
in which the second-order low-pass filter $ M^{\sfLP}$ has unity gain at low frequencies, with the exact  pole/zero locations used for the pool loads.   The scaling of $1/4$ is introduced so that the response from the ideal low frequency actuators is commensurate with the pools.
The resulting Bode plot is no longer flat -- its gain below $10^{-4}$~rads/sec is approximately half of the gain above $10^{-3}$~rads/sec.

%      \begin{figure}[h]
%      	\Ebox{.75}{OpenLoopAggregate.pdf}
%      	\vspace{-1em}
%      	\caption{\textit{Open-loop tracking} with residential air-conditioners,
%      		electric water heaters, pool pumps, and ideal actuators. Low frequency error is a result of insufficient resources in lower frequency bands.}
%      	\vspace{-.3em}
%      	\label{f:trackingIII}   	\vspace{-.3em}
%      \end{figure}
%      
%
%   
%    
%   
%\Fig{f:trackingIII} compares the BRD with the total response from all of these actuators. 

% Joel starts : do we need this paragraph as we have deleted the figure
As a result of the gain variations in the linearization and the nonlinearities caused by capacity constraints, the open-loop tracking will no longer be perfect, especially when the BRD signal takes on large values.    While imperfect, the performance is still better than what is received from many generation units (such as  Fig.~10 of \cite{kir04}).    
% Joel ends
\spm{Figure removed due to lack of space.  It also isn't very informative}

   \begin{figure*}
   	\Ebox{1}{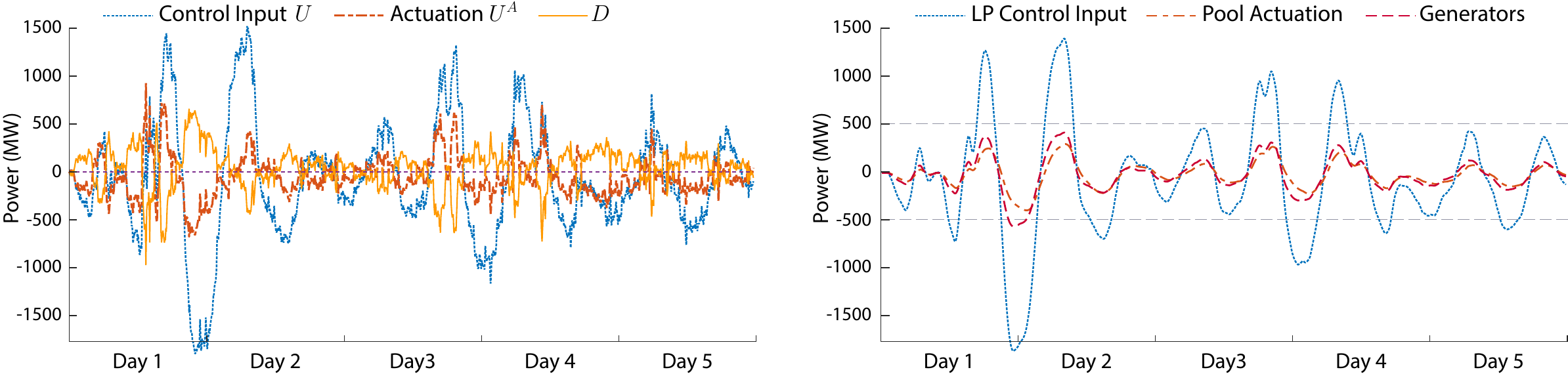}
   	\vspace{-1em}
   	\caption{Closed-loop tracking with residential air-conditioners,
	electric water heaters, pool pumps, and ideal actuators. Actuation from the loads can be interpreted as virtual energy storage.} 
   	\vspace{-.3em}
   	\label{f:ClosedLoopOverallMillionPools}
   \end{figure*}
   
The following set of experiments are based on the closed-loop architecture used in practice today: the BA  observes frequency deviations (or some other measure of power mismatch),  and varies the balancing reserve/AGC signal in response.  Results from these experiments are described next.
%
%\spm{need to explain AGC vs BRD somewhere}

The plots on the left hand side of \Fig{f:ClosedLoopOverallMillionPools} show the resulting  closed loop behavior over 5 days, using BPA BRD data from February 19--23, 2016, as the disturbance $\bfmD$ entering the grid (modeled as an additive input disturbance as shown in  \Fig{f:grid}).

\spm{This isn't the place for a general philosophical discussion.   The origins of $\bfmD$ belong in a general model
description early in the paper.  In a numerics section we want PRECISION:   what was done, why was it done this way, and what are the conclusions.
\\
; in general, continuous disturbances can include load forecast errors, renewable generation forecast errors, etc.
}
The aggregate response from all actuators, $\bfmU^A$, is approximately the negative of the BRD, so that the frequency deviation is tightly controlled:  the disturbance rejection performance  is nearly perfect.
 The grid frequency remained within the range  59.993 to 60.007 Hz over the 5 day period.

The plot on the right hand side of \Fig{f:ClosedLoopOverallMillionPools} shows the filtered control signal   $U^{\sfLP}_t\eqdef M^{\sfLP} U_t$ along with two  responses:  from the collection of pools, and from the 1~GW generator. 
 The response of the pools nearly matches the response from the ideal generator.

% (i) For disturbance = BPA BRD : 59.993 to 60.007 hz
%(ii) For disturbance = CAISO ACE: 59.98 to 60.02 hz

%\rd{................}
%
%
%The response of the system to sudden shocks is also a question of interest. To study this, we observe the closed loop behavior to the following disturbance: a 3 GW generator goes offline for 30 mins. The closed loop performance is shown in \rd{f:IIIB2}. When sufficient number of loads are present in the demand dispatch model, the system can reject shocks without any impact on QoS of the loads; in other words, loads participating in the demand dispatch model also serve as contingency reserves. While perturbations can be observed in the control input and the output of the actuators for a few hours after the event, the disturbance rejection performance is nearly perfect. The capacity metrics of the loads will be studied in future work.

\begin{figure}[h]
	\Ebox{.6}{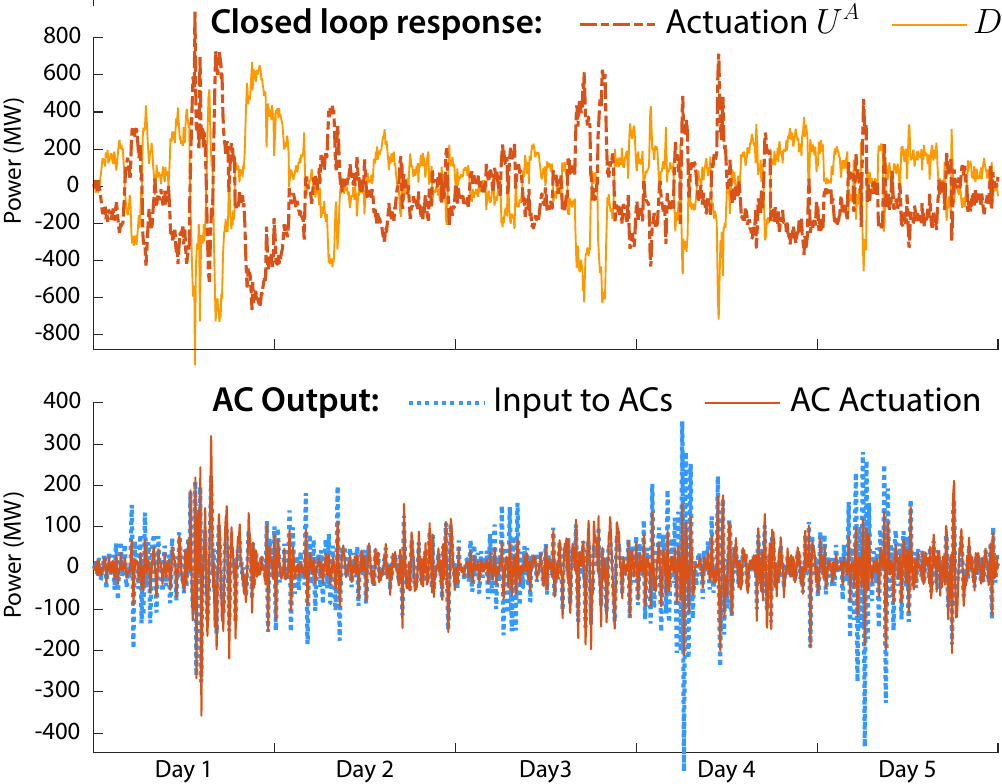}
	\vspace{-1 em}
	\caption{Tracking remains perfect even when the gain from ACs is sinusoidal with a 24 hour period and   magnitude range $1\pm \half$.}  % $[1/2,3/2]$.} 
	\vspace{-.3em}
	\label{f:periodicAC}
\end{figure}

\subsection{Time-varying capacity}
\label{s:TimeVarying}

The time-varying nature of many commercial and residential loads is an issue of concern. For example, the number of air conditioners that are in operation, and hence available for ancillary service, is low during the early morning hours and peaks during the late afternoon --- see Fig. 9. of \cite{smimendonsim13}.

Experiments were conducted in which the gain of the response of the ACs was amplified/attenuated using a time-varying gain  function:
\[
g(t) = 1 - 0.5 \sin (f_d t ),\qquad t\ge 0
\]
in which $f_d=727\times 10^{-7} $~rads/s corresponds to  a 24 hour period.   
All of the other resources were left the same as the simulation setting of \Section{s:cl}.
\Fig{f:periodicAC} shows the results. While the AC actuation does not track its input signal $\bfzeta^\sfac$, the aggregate actuation from all the resources is  almost the exact opposite of the disturbance, just as seen in previous experiments.  Disturbance rejection is nearly perfect, and the grid frequency remains within  [59.993, 60.007] Hz.

The potential cost of these gain fluctuation is additional actuation from other resources  \cite{matkadbusmey16}.

\subsection{Resource availability and  cost}
\label{s:FastCost}

Following installation of equipment to enable demand dispatch,  the operating cost is essentially zero.   Consumers may require incentives to participate (e.g., the monthly credits provided by Florida Power and Light through their OnCall program),  but they will also receive some guarantees regarding constraints on QoS and potential costs from additional cycling of equipment.   

The benefit of demand dispatch from low frequency services such as residential pools is clear:   one million pools serve as a substitute for a 1GW generator.  Following the initial investment (usually in \$B), a generator requires fuel, maintenance, and   staff.  The   loads provide accurate regulation service without any of these operating costs.

\begin{figure}[h]
	\Ebox{.6}{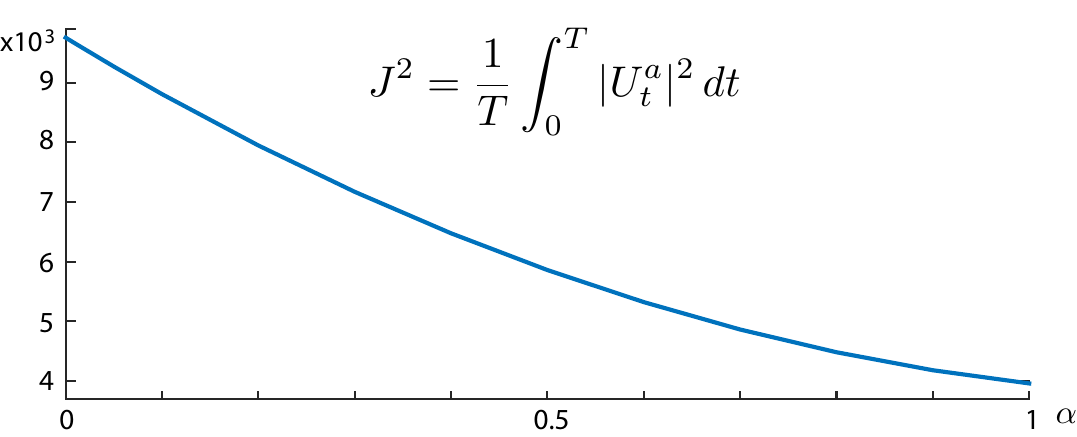}
	\vspace{-1 em}
	\caption{Cost as a function of capacity from AC loads.} 
	\vspace{-.3em}
	\label{f:cost}
		\vspace{-.3em}
\end{figure}

What about high-frequency ancillary services?  To investigate the value of the highest frequency services from demand dispatch, we consider a parameterized family of models in which the contribution from air conditioners is varied according to the fraction $\alpha \in [0,1]$.   The remaining $1-\alpha$ of regulation is obtained from ideal actuation from batteries or other sources.     Denote the output of the ideal actuators by $\{U^a_t\}$.  
The total ideal actuation is defined by the sum: 
\[
U^a_t =  [M^{\sfHP} G_a + (1-\alpha) M^{\sfBP}_{\sfac}  G_a] U_t
\] 
Recall $G_a\equiv 1$ in these experiments.  The second component  is thus $(1-\alpha) M^{\sfBP}_{\sfac}  U_t$,   which is intended to replace the lost service from the ACs.

\begin{figure*}
	\Ebox{1}{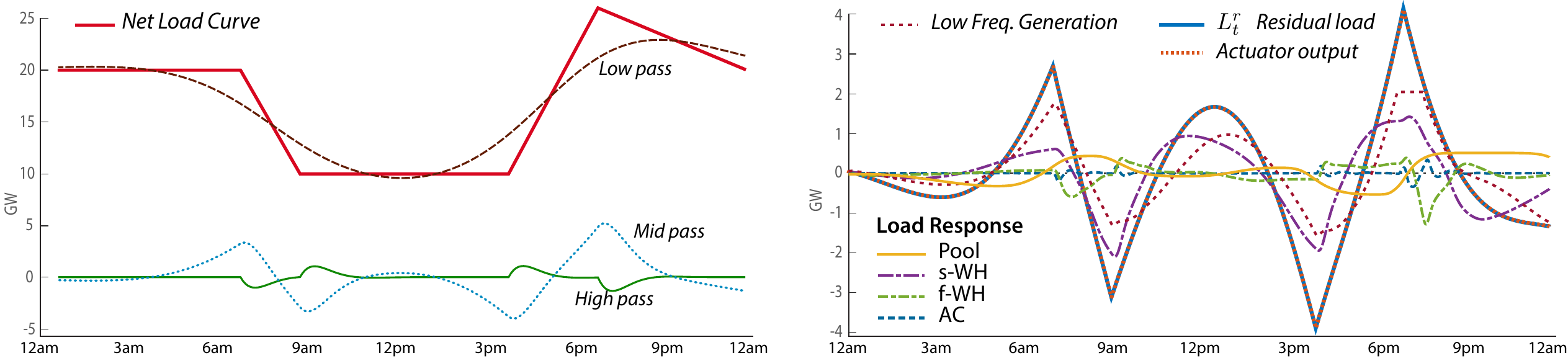}
	\vspace{-1 em}
	\caption{Left: Hypothetical CAISO Net-load over one day in 2020, and its frequency decomposition.
	Right:  {\sl ``Residual Load'' $=$ ``Net Load''$-$``Low Pass''} is
	 tracked nearly perfectly.   The introduction
	of demand dispatch alongside generation reduces needed generation capacity by at least 5~GW.} 
	\vspace{-.3em}
	\label{f:duck}
\end{figure*}

 Following \cite{matkadbusmey16}, the mean-square cost of the closed loop system is defined as,
\begin{equation}
J^2 =  \frac{1}{T} \int_0^T | U^a_t  |^2\, dt  %\lim_{T \to \infty}
\end{equation}
This is similar to the ``mileage'' metric used for ancillary service resources such as batteries.
\Fig{f:cost} shows a plot of this cost as a function of $\alpha$ for $T$ corresponding to one day;    $\alpha=1$ corresponds to the simulation setting of \Section{s:cl}. The total cost is reduced by more than 50\%\ when $\alpha=1$ as compared to $\alpha=0$.

 The cost would be much higher for intermediate values of $\alpha$ if the inverse filter was not used to construct  $\bfzeta^\sfac$   \cite{matkadbusmey16}.

%T = one day (86,400 s). \alpha = 1 implies 1 million ACs. I am not sure about the initial steep decline; i checked for \alpha = 0 and then the next increment was \alpha = 0.02. I can check for \alpha between these values if we need it.

%\rd{Lots of questions here.  First,   {$T=$ on day?}
%\\
% Second,
%we should have a list of bandwidths earlier and not repeat here.   Second,  $H_\alpha$ refers to a linearization.  I don't think it is useful here, but I may be misunderstanding something.
%\\
%The setup was as follows. For $ \alpha \in [0,1]$, the proportion was ACs and ideal actuators servicing the frequencies from 0.002 to 0.008 rad/s varies as,
%\[
%H_{\alpha} = M^{\sfBP}_{\sfac} [ \alpha M^{\sfinv}_{\sfac} G_{\sfac} + (1-\alpha) G_a ]
%\]
%where $M^{\sfBP}_{\sfac}$ and $M^{\sfinv}_{\sfac}$ are the local bandpass and inverse filters for ACs, respectively. 
%}

\subsection{Ramp services}
\label{s:IIIramps}

The plot on the right in \Fig{f:duck} shows a stylized ``duck curve'' representing the net-load at CAISO anticipated in the near future, based on the assumption that there will be significant solar energy penetration.  The plot is based on  approximately 10 GW of solar power at peak.   
\spm{The figure is taken from our 2016 HICSS presentation.}

%
%\begin{figure}[h]
%	\Ebox{.75}{FilteredDuckHICSS2017.pdf}
%	\vspace{-1 em}
%	\caption{Hypothetical CAISO Net-load over one day in 2020, and its frequency decomposition.} 
%	\vspace{-.3em}
%	\label{f:duck}
%\end{figure}

The 15GW ramp observed between 3pm and 6pm is of concern today.    It is argued in  \cite{matkadbusmey16} that the ramp can be smoothed by first scheduling generation to track a low-frequency component of the net-load --- denoted ``low pass'' in the figure.   The remaining two zero-energy signals shown can be tracked using a combination of resources --- batteries, responsive generators, and demand dispatch.

The mid-pass signal remains substantial -- a range of $\pm$~5GW.      
This signal could be provided using  gas turbine generators, but a total capacity of 10GW would be required.  This value can be reduced significantly by applying the same techniques used to address the balancing reserves signal.

Let $L^r_t $ denote the residual load, defined as ``Net Load''$-$``Low Pass''.   This is plotted on the right in \Fig{f:duck}, where it is seen that it takes on values approaching $\pm$~4GW.   The capacity from loads in the previous set of experiments was insufficient to track this signal.     The capacity from TCLs was doubled, so that the simulation was based on  10 million s-WH, 
 10 million f-WH, and 2 million ACs.   It included  1.2 million pools (the approximate number of pools in California),  and also $\pm$~2GW of low frequency regulation that might come from generation or demand dispatch from other loads such as water chillers and water pumping (a significant load in California).

The plots of deviation of power from TCLs shown on the right in \Fig{f:duck} are significant, even though the loads themselves do not deviate from their individual temperature setpoints.   The variation in power consumption of s-WH and pools helps to address the ``mid pass''  signal shown on the left of \Fig{f:duck}, whereas the ``high pass''  component is serviced by the f-WH and AC power consumption.  The residual load and aggregate actuation match nearly perfectly.

%\begin{figure}[h]
%	\Ebox{.75}{ManagingTheDuck3flippedLoads.pdf}
%	\vspace{-1 em}
%	\caption{{\sl ``Residual Load'' $=$ ``Net Load''$-$``Low Pass''} is
%	 tracked nearly perfectly.   The introduction
%	of demand dispatch alongside generation reduces needed generation capacity by at least 5~GW.   } 
%	\vspace{-.3em}
%	\label{f:ManagingTheDuck2}
%	\vspace{-.3em}
%\end{figure}

\section{Conclusions}
\label{s:conc}

It is exciting to see how  heterogeneous loads can coordinate through distributed control to smooth out enormous shocks to the grid. The collection of heterogeneous loads is a multi-GW virtual battery capable of impressive actuation in response to the control signal from the grid operator. In a closed-loop setting, the demand dispatch architecture can perform near-perfect disturbance rejection, tightly controlling the grid frequency.
\spm{too US centric!  in close proximity of 60 Hz. }
Consequently, demand dispatch offers tremendous potential to provide high-quality ancillary services on timescales spanning from several hours to a few minutes (the time-scale of AGC). 

\spm{This was only true for very low-capacity service.  I'm sure it is much greater in the macro experiments:
In addition, experiments in this paper indicate a less than 5\%\ deviation from normal cycling for a typical TCL participating in demand dispatch.}

\spm{Ana objects: \textit{What have we missed?}  
}

Two issues require further attention.  First is the role of the ``perfect actuators'' supplying regulation at time scales of tens of seconds and faster (the timescale of today's primary control).   Can loads assist with this service as well as bolster synthetic inertia?   The analysis in \cite{matkadbusmey16} suggests that this could bring risk in terms of stability,  but this may depend on other elements of the grid architecture (e.g.,  the number and size of  synchronous generators).

A second, far more significant issue is the time-varying nature of many loads.   For example,  the nominal load from commercial and residential air-conditioning is roughly periodic over a typical week, and its magnitude changes slowly depending upon the weather. The results summarized in \Section{s:TimeVarying} offer significant hope in terms of system stability.   Moreover, it   is conjectured that  periodicity is a benefit in regions with significant solar energy, since demand is in harmony with supply.  

Future work is required to convince the scientific community and the power industry that the overall coupled dynamics will not introduce any additional risk when compared to traditional methods for balancing and frequency regulation.  Further large-scale simulation is required along with large-scale demonstration projects.

   \spm{I am suspicious about the page numbers for
    {matdyscalros12}: 1000(2000):3000, 2012.
    \\
    It seems the pages are actually
    189-203}
 
%\bigskip
\medskip

%\bibliography{strings,markov,q,extras}

\bibliographystyle{abbrv}  %abbrv

%  {\small
 %}

\def\cprime{$'$}\def\cprime{$'$}

%\null  %Needed with \usepackage{flushend}

%\clearpage
\bigskip

\appendix

\section{Appendix}

\subsection{Grid-level transfer functions}

	\Table{t:TF} provides the transfer functions of the macro grid model $G_p$ and the PI compensator $G_c$ used in the closed-loop experiments of \Section{s:sim}.
	
	\begin{table}[h]
\centering		
	\begin{tabular}{||c c||}  
		LTI System & Transfer Function
		\\ [2ex]  
		$G_p$ &  $\displaystyle 10^{-5} \,\frac{ 2.488 s + 2.057}{s^2 + 0.3827  s + 0.1071}$
		\\ [3ex]
		$G_c$ & $\displaystyle 516\, \frac{s + 0.5}{s}$
	\end{tabular}
	\caption{Continuous-time transfer functions used in the closed-loop simulations.}
	\label{t:TF}
\end{table}

\subsection{Optimal inverse filter design}

Consider the feedback system shown in \Fig{f:localFilterDesign}.  Recall that the associated \textit{loop transfer function} is defined as the product $L_\ell =K_\ell G_\ell $,  and the closed-loop transfer function   is expressed,
\begin{equation*}
\label{e:T}
T_\ell  = \frac{L_\ell }{1+L_\ell }
\end{equation*}
The transfer function  $K_\ell $ is designed so that $T_\ell $ can be approximated by a band pass filter with given bandwidth, denoted $\Omega_{des}$:
\begin{equation}
\begin{aligned}
T_\ell (j\omega) \approx 1, \qquad \omega\in \Omega_{des} \\
and \quad T_\ell (j\omega) \ll 1, \qquad \omega \notin \Omega_{des}
\end{aligned}
\label{e:Tbound}
\end{equation}   
The control theory literature has many tools for successful design of $K_\ell$ to achieve this goal.

Once we have managed to achieve \eqref{e:Tbound} through choice of $K_\ell$,  we obtain the desired inverse filter via
\begin{equation}
\label{e:M}
M_\ell  = \frac{K_\ell }{1+L_\ell }
\end{equation}
The resulting transfer function from input $u_t$ to output $y_t$ is thus,
\begin{equation}
H_\ell   \eqdef M_\ell  G_\ell  = T_\ell  
\label{e:H}
\end{equation}

We now give details on one approach to design $K_\ell$.
The transfer function $T_\ell $ is   known as the \textit{complementary sensitivity function}.  The \textit{sensitivity function} is 
\begin{equation*}
\label{e:S}
S_\ell  \eqdef 1-T_\ell = \frac{1}{1+L_\ell }
\end{equation*}
Our goal is to choose 
$K_\ell $ such that $T_\ell (j\omega)\approx 0$ for $\omega\notin\Omega_{des}$,  $S_\ell (j\omega)\approx 0$ for $\omega\in \Omega_{des}$,  while maintaining reasonable bounds on $|M_\ell (j\omega)|$ for all $\omega$.

We utilize the mixed-sensitivity synthesis method to construct $K_\ell$  \cite{kwa93}.
This requires three    transfer functions  $(W_1, W_2, W_3)$ that serve as weights for the respective transfer functions $(S_\ell,  M_\ell, T_\ell)$.  For any transfer function $K$ we obtain a three-dimensional transfer function $ Q = ( W_1 S_\ell , W_2 M_\ell ,W_3T_\ell)$,  where for example, when $W_1 = 1$, $Q_1 = S_\ell =1/(1+K G_\ell)$.  The   $H_\infty$-norm of $Q$ is denoted
\[
q(K) = \| Q\|_\infty\eqdef \max_i \|Q_i \|_\infty 
\]
where the individual  norms are given by,
\[
\|Q_1\|_\infty = 
\max_\omega | W_1(j\omega) S_\ell (j\omega) | ,
\quad
\|Q_2\|_\infty =\max_\omega |W_2(j\omega) M_\ell(j\omega)| ,
\quad
\|Q_3\|_\infty = \max_\omega |W_3(j\omega)T_\ell(j\omega)|
\]
The mixed-sensitivity synthesis method finds the transfer function that minimizes $q(K)$ 
over all proper transfer functions $K$.  We define $K_\ell$ to be this optimizer:  
\begin{equation}
\begin{aligned}
K_\ell&= \argmin_K q(K)  
\end{aligned}
\label{e:Kmin}
\end{equation} 
This can be solved numerically, using the {\it mixsyn} command in MATLAB \cite{bal05}.

\subsection{Open loop simulations of TCLs}

The open loop simulations in this paper involve three groups of TCLs: ACs, f-wh, and s-wh. Each group consists of 20 sub-groups containing 2,000 similar loads in each subgroup. 

The temperature evolution of an individual load is defined by the ODE given in \eqref{e:tcltemp}. The different subgroups are obtained by uniformly sampling the TCL parameters provided in \Table{table:tclparameters}.

For a given TCL, a finite state space for $\staten$ is obtained through the quantization of the interval $[\theta_{min},\theta_{max}]$, where $\theta_{min} = \theta_{set} - \delta/2$ and $\theta_{max} = \theta_{set} + \delta/2$. For a given integer $d$, the interval $[\theta_{min}, \theta_{max}]$ is discretized into $d/2$ values as follows: 
\begin{equation*}
\staten = \{ \theta_{min} + k \theta_{\Delta} : 0 \leq k \leq d/2-1 \}
\end{equation*}
where $\theta_{\Delta} = (\theta_{max} - \theta_{min})/(d/2 - 1)$ represents the temperature increments in the interval $[\theta_{min}, \theta_{max}]$. Furthermore, $\stateu(t) = m_t \in \{0,1\} \equiv \{\ominus,\oplus \} $.

The matrix $Q_0$ modeling transitions in $\staten$ is obtained via Monte Carlo simulations of \eqref{e:tcltemp}. 

 Let $\{T_k: (k \geq 1)\}$ denote the jump times in a Poisson process with rate $r$.  Let $\{ \Delta_k \eqdef T_k - T_{k-1} : k \geq 1\}$ and $T_0 = 0$. It follows that $\Delta_k$ is i.i.d., with $\mathrm{E[\Delta_k] = 1/r}$. Then, $Q_0$ is identified via Monte Carlo methods as below.
 
 The bivariate distribution from a given state $x$ to a particular temperature state $x'_n$ is obtained as an empirical average,
 \begin{equation*}
 \pi_{Q}(x,x'_n) = \frac{1}{N}\sum_{k=0}^{N-1} \ind(X(T_k) = x, X^n(T_{k+1}) = x'_n),
 \end{equation*}
 where  $x \in \state, x'_n \in \staten$.   Bayes rule motivates the following definition for the transition matrix:
 \begin{equation*}
 \label{e:MCgeo}
 \begin{split}
 Q_0(x,x'_n) &= \Prob\{X^n(T_{k+1}) = x'_n | X(T_k) = x\} \\
 &= \frac{\pi_{Q}(x,x'_n)}{\sum_{x''_n}\pi_{Q}(x,x''_n)}
 \end{split}
 \end{equation*}

Following the notation established in \cite{meybarbusyueehr15}, $p^\oplus(x_n)$ and $p^\ominus(x_n)$ indicate the probability of a TCL unit switching on and switching off, respectively, at temperature state $x_n$. The transition matrix $R_0$  that models the on/off behavior (i.e. transitions in $\stateu$) is represented as,
\begin{align}
\label{e:R0}
\begin{split}
R_0(x,\oplus) = 
\begin{cases}
1 - p^\ominus(x_n) \quad &x = (\oplus, x_n) \\
p^\oplus(x_n) \quad &x = (\ominus, x_n)
\end{cases} \\
R_0(x,\ominus) = 
\begin{cases}
p^\ominus(x_n) \quad &x = (\oplus, x_n) \\
1 - p^\oplus(x_n) \quad &x = (\ominus, x_n)
\end{cases}
\end{split}
\end{align}

\begin{figure}[!h]
	\Ebox{.75}{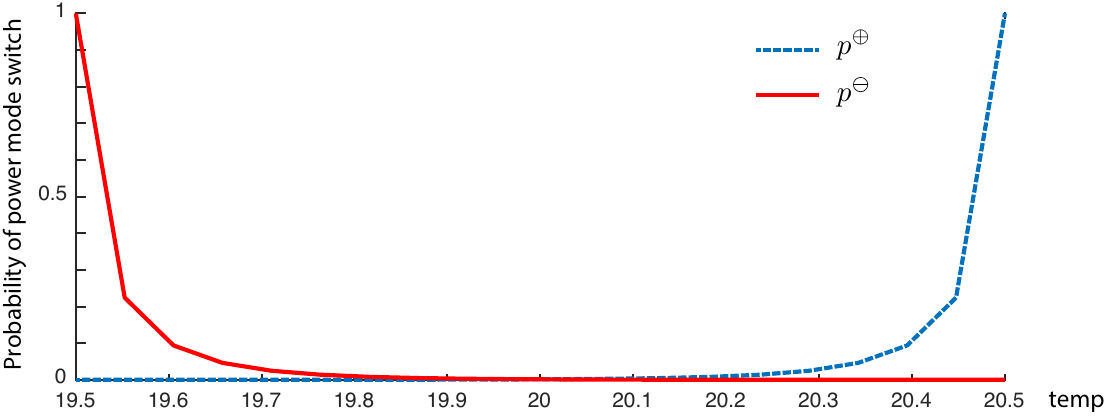}
	\vspace{-.25em}
	\caption{Probabilities defining $R_0$ for an AC.} 
	\label{f:switchprob1} 
\end{figure}

\Fig{f:switchprob1} shows a particular choice of $\{p^\ominus,p^\oplus\}$ for a given AC, in which $p^\oplus(x_n) = p^\ominus(\theta_{max} + \theta_{min} - x_n)$ for each $x_n \in \staten$. 

The Markov transition matrix $S_0$ can be obtained from \eqref{e:PQ0R0}, and the rate matrix $\generate_0$ can be generated using \eqref{e:rate}.

In this paper, the TCL temperature interval is quantized into $d/2 = 20$ temperature states. Consequently, $|\state| = d = 40$.  

To model the behavior shown in \Fig{f:tempvtime1}, we set $r=1/60~s^{-1}$ for ACs. Fast water heaters (f-wh) are on for $15$ minutes and off for $2.5$ hours, while slow water heaters (s-wh) remain on for $25$ minutes and off for $9$  hours, approximately. To model this asymmetry in on-off durations, we utilize two rate parameters: $r_{\mathrm{on}}$ and $r_{\mathrm{off}}$. We set $r_{\mathrm{on}} = 1/40 ~s^{-1}$  and $r_{\mathrm{off}} = 1/500~s^{-1}$ for f-wh, and $r_{\mathrm{on}} = 1/40 ~s^{-1}$  and $r_{\mathrm{off}} = 1/1000 ~s^{-1}$ for s-wh.

The BPA balancing reserves (from February 19, 2016) are used as the control signal $U$. The open-loop simulations are in discrete-time; the TCLs receive $U$ at 20-second intervals. $U$ is passed through a local load-level second-order Butterworth filter; the passband frequency range of this filter is specified in \Table{t:loadchar}. An additional inverse filter is used, as discussed in \Section{s:LC:Inv}, in order to obtain a flat input-output response from the aggregate collection of loads in each sub-group; this yields the signal $\zeta$.

The matrix $S_{\zeta}$ is obtained using the myopic design defined in \eqref{e:PQ0R0myopic}. The mean-field behavior of each group of loads is realized via \eqref{e:ctsMFM}. The linear models are specified in \eqref{e:LSSmfg}. Note that the Markov transition matrix for a given discrete-time setting $t$ can be obtained from the rate matrix as $P_{\zeta}^t = \exp(t \generate_{\zeta})$,  $t\ge 0$, where $P_{\zeta}^t(x,x') = \Prob\{X_t=x'\mid X_0=x\}$ is the transition probability from state $x$ to $x'$, with $x, x' \in \state$. The stochastic behavior of the loads specified by the transition matrix, i.e. the transition from the current state $x$ to the next state $x'$ for a given load, can be implemented using a uniformly distributed random number generator $\sim \mathcal{U}[0,1]$.

\subsection{Closed loop simulations}

\begin{figure}
	\Ebox{1}{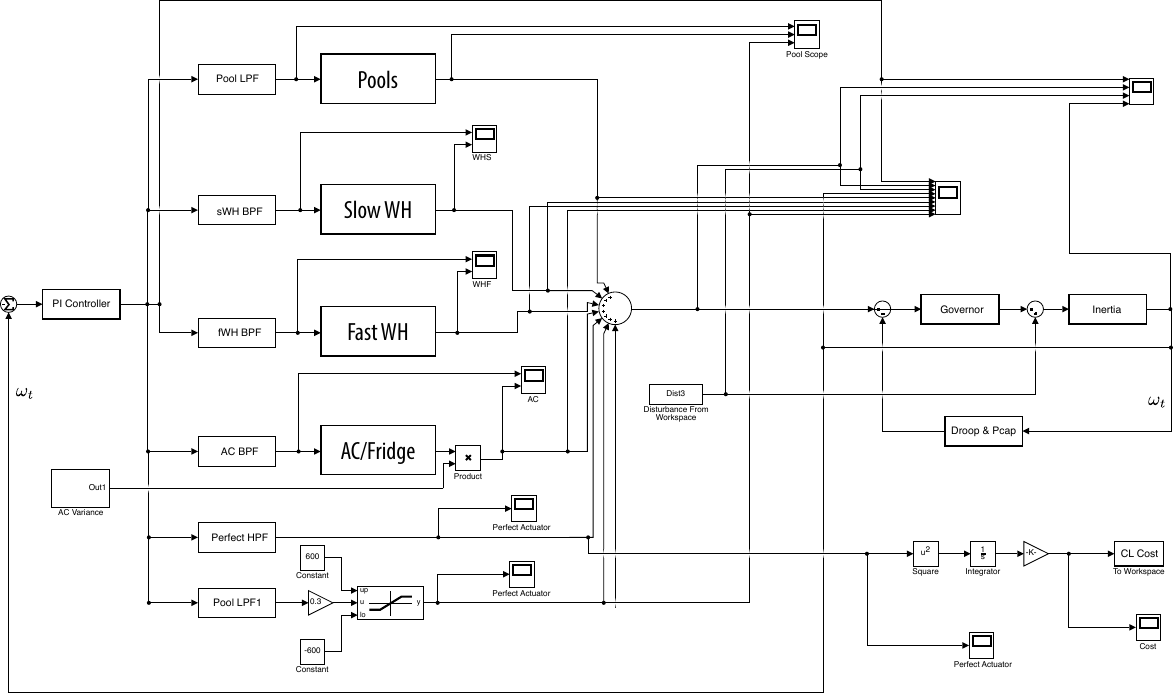}
	\vspace{-1 em}
	\caption{Simulink model.} 
	\vspace{-.75em}
	\label{f:SimulinkGridV3}
\end{figure}

The closed loop system shown in \Fig{f:grid} and discussed in \Section{s:cl} is implemented in continuous time using Simulink.  The full Simulink model is illustrated in 	\Fig{f:SimulinkGridV3}.
 \Fig{f:fig1ERCOT-Grid} is taken from Fig.~1 of \cite{chabalsha12}, on which this Simulink model is based.

 Data from the BPA balancing reserves (February 19--23, 2016) is used for the disturbance $D$.

The nominal models ($S_0, \generate_0$) of the demand-side resources, which include ACs, f-wh, s-wh, and pools,  are generated using open loop simulations (as described in the paper on); the mean-field models are defined by the nonlinear deterministic equation \eqref{e:ctsMFM}, which is non-linear in $\zeta$. The mean-field models are implemented using the Linear Parameter Varying (LPV) block in Simulink, with  $\zeta$ as the parameter.  

The bandpass and inverse filters are implemented as continuous-time linear systems and are similar to the ones used in the open-loop simulations; the bandpass filters are second-order Butterworth filters with frequency ranges provided in \Table{t:loadchar}. The grid-level transfer functions $G_p$ and $G_c$ given in \Table{t:TF} are also implemented using Simulink's continuous-time LTI system block.

  \end{document}